\documentclass[english]{article}
\usepackage[latin9]{inputenc}
\usepackage{amssymb}
\usepackage{esint}

\makeatletter
\date{}

\makeatother

\usepackage{babel}
\begin{document}
\title{Comments on the nonlinear Schrödinger equation}
\author{Mark Davidson\thanks{Spectel Research Corp., Palo Alto, CA, USA; mdavid@spectelresearch.com} }
\maketitle
\begin{abstract}
A proof is given that if the nonlinear Schrödinger wave function is
constrained to have support over only a finite volume in configuration
space, then the total energy is bounded from below for either sign
of the logarithmic term in the Hamiltonian. It is concluded that the
usual assumption about the sign of the logarithmic term made by Bialynicki-Birula
and Mycielski is not the only possibility, and that a sensible theory
can be made with the opposite sign as well.


\noindent \textbf{Journal}: Il Nuovo Cimento B, V116B, 1291-1296 (2001).
ISSN 0369-3554
\end{abstract}

\section{Introduction}

A generalization of Schrödinger\textquoteright s equation to include
a nonlinear logarithmic potential was first proposed by Bialynicki-Birula
and Mycielski {[}1{]}. In this work, only one sign of the logarithmic
term was considered, as the other sign led to a Hamiltonian which
was not bounded from below. Soliton solutions were exhibited in {[}1{]},
and it was proved that the Hamiltonian was bounded from below for
one sign choice.

A logarithmic term was also proposed long ago within the context of
stochastic quantum mechanics {[}2,3{]}. In {[}2{]} the logarithmic
term could have either sign. In {[}3{]} a speculative stochastic-electromagnetic
model of Schrödinger's equation was proposed which led to the existence
of the logarithmic term with a positive sign (opposite in sign to
{[}1{]}), which could be interpreted as a diffusion force. Weinberg
{[}4{]} also considered the implications of a general class of nonlinear
Schrödinger equations including the logarithmic one. 

The nonlinear Schrodinger equation considered here is:

\begin{equation}
\left[-\frac{\hslash^{2}}{2m}\triangle+V+kTln(\psi^{*}\psi v_{0})\right]\psi=i\hslash\frac{\partial\psi}{\partial t}
\end{equation}

\noindent Here k is Boltzman's constant, T has units of temperature,
and $v_{0}$ is an arbitrary constant volume to make the dimensions
of he logarithm argument dimensionless. Only the single particle equation
is considered, as the generalization to other Hamiltonians is straightforward.
Note that if we add a multiplicative constant term inside the logarithm,
it has no physical effect as it just adds a constant to the Hamiltonian.
Thus changing $v_{0}$ just adds a different constant to the Hamiltonian.
To conform with {[}1{]} T would be negative, and to conform with {[}2,3{]}
it could be positive. 

There has been much theoretical and experimental analysis looking
for evidence of a nonlinear term. Lamb shift calculations {[}1{]}
led to a limit of $\left|kT\right|<4\times10^{-10}eV$ for it. Shimony
{[}5{]} proposed an experiment using coherent thermal neutron interferometry.
These experiments were performed by Shull et al.{[}6,7{]} with the
result that $\left|kT\right|<3.4\times10^{-13}eV$ , improving the
bound by three orders of magnitude. Gähler et al. {[}8{]} measured
the Fresnel diffraction of coherent thermal neutrons at a sharp edge
and were able to lower the bound to $\left|kT\right|<3.3\times10^{-15}eV$.
The theoretical analysis used in these coherent neutron papers ignored
the combined temporal and spatial incoherence of the thermal neutron
beam, and so their validity might be questioned considering the nonlinear
interaction term. This shortcoming was pointed out in {[}8{]}. However,
an analysis of the fully incoherent case which is not presented here
for brevity has concluded that the coherent approximations and the
results claimed in {[}5-8{]} appear to be justified. Therefore, it
seems that the logarithmic term, if it exists at all, is rather small.
It is important nevertheless to continue the search for a nonlinear
term, because any such term would help resolve the deep conceptual
problems which the various interpretations of quantum theory pose.
If Schrödinger's equation is exactly linear then this is hard to explain
in a stochastic model of quantum mechanics where extra terms tend
to appear.

Gisin {[}9,10{]} has further argued that if there is a nonlinear term
of any kind in Schrödinger's equation, then signals could in principle
be sent faster than the speed of light, leading to a causality dilemma.
Gisin's argument seems sound, but the subject is somewhat challenging
conceptually, and it is possible that some modification in Gisin's
assumptions might allow a logarithmic nonlinearity to peacefully coexist
with causality. 

The logarithmic nonlinearity has a number of unique properties that
make it the best candidate for a nonlinear correction to Schrödinger's
equation aside from the causality issue {[}2{]}. 

Property 1 - The total integrated force caused by the logarithmic
term vanishes.

Property 2 - The total integrated torque about any center caused by
the logarithmic term also vanishes. 

Property 3 - If a wave function satisfies (1) then any constant times
that wave function also yields a solution. The normalization adjustment
simply shifts the logarithmic term by a constant energy factor which
can be ignored.

Property 4 - The logarithmic perturbation preserves factorization
properties since the log of a product is a sum of logs. So factorizing
solutions can still be found for Coulomb potentials in spherical coordinates
for example. Also, in multiparticle systems, it allows the particles
to be independent of one another.

Property 5 - A plane wave is still an exact solution to the wave equation,
although linear superposition no longer holds.

Properties 1 and 2 are necessary in order to agree with the classical
correspondence limit. If the nonlinear term led to a net force or
torque this would be apparent with large objects. Properties 3 and
4 are desirable for a probability interpretation.

\section{The Hamiltonian Bounds}

The quantum average of the Hamiltonian is

\begin{equation}
H=\intop\psi^{*}\left[-\frac{\hslash^{2}}{2m}\triangle+V+kTln(\psi^{*}\psi v_{0})\right]\psi d^{3}x
\end{equation}

\noindent Bialynicki-Birula and Mycielski {[}1{]} reject the positive
sign for T because the integrated Hamiltonian for (1) is not bounded
from below in this case. To see why this is, consider a wave function
which is very constant and slowly varying in the absence of any external
force. Then the kinetic term is insignificant and we have that 

\begin{equation}
H=kT\int\rho ln(\rho v_{0})d^{3}x,\qquad\rho=\psi^{*}\psi
\end{equation}

\noindent This expression is clearly not bounded from below for positive
T as we see below. The integrand in (3) is bounded from below however
since we have

\begin{equation}
f(\rho)\equiv\rho ln(\rho v_{0}),\:\frac{df}{d\rho}=1+ln(\rho v_{0})=0\:at\:the\:extremum
\end{equation}

\noindent and therefore

\begin{equation}
\rho ln(\rho v_{0})\geq-\frac{1}{ev_{0}}
\end{equation}

\noindent This must be true at the extremum, no matter what units
we had chosen. Now suppose that the wave function is constrained to
have support over only a finite volume. This volume could be large,
say the size of the observable universe. No matter how large it is,
the Hamiltonian is then bounded from below by:

\begin{equation}
H\geq-kT\frac{volume}{ev_{0}}
\end{equation}

\noindent A stronger and better bound can be derived since $\rho$
satisfies a normalization condition

\begin{equation}
\intop_{volume}\rho d^{3}x=1
\end{equation}

\noindent We can use the calculus of variations to minimize $f$ subject
to the constraint (7). To achieve this we use a Lagrange multiplier
$\lambda$. Let S denote the functional to be rendered extremal.

\begin{equation}
S=\intop_{volume}\left[\rho ln(\rho v_{o})+\lambda\left(\rho-\frac{1}{volume}\right)\right]d^{3}x
\end{equation}

\noindent The solution to the extremum problem is simply that $\rho$
is independent of x and therefore $\rho=1/volume$ and so 

\begin{equation}
H\geq-kTln(\frac{volume}{v_{0}})
\end{equation}

\noindent So we see from (9) that if the volume tends to infinity,
the Hamiltonian is not bounded from below, but it is bounded as long
as the volume is constrained to be a finite value. 

The requirement that the wave function be constrained to have support
over at most only a fixed finite volume is a very weak one. Certainly
if wave functions are allowed to have support only over the observable
universe then this would not have any adverse consequences for any
quantum mechanical calculations. No experimental predictions of quantum
mechanics would be in any way be affected by such a constraint. And
therefore for all practical purposes the Hamiltonian is bounded from
below for both signs of T, and the positive sign for the logarithmic
term should not be dismissed out of hand as has been done up till
now. The negative sign was shown in {[}1{]} to have soliton solutions.
The positive sign for the logarithmic term does not have soliton solutions.
It causes a slightly faster spreading of wave packets than quantum
mechanics would predict. It has the interpretation of a diffusion
force.

\section{Other Nonlinear Models}

Other nonlinear forms have been proposed within the context of stochastic
quantum mechanics such as the stochastic electrodynamic calculation
in {[}11{]} where a phase space model based on the the theory of electron
diffusion in gases is presented and applied to the problem of an electron
subject to classical zero point radiation. There it was argued that
the Schrödinger equation is a good approximation to the diffusion
equations. The leading correction term in the Hamiltonian has the
form

\begin{equation}
H_{1}(\rho)=-b\left[\frac{1}{\rho}\nabla^{2}\nabla^{2}\rho+\nabla^{2}\left[\frac{1}{\rho}\nabla^{2}\rho-\frac{1}{\rho^{2}}\left(\nabla\rho\right)^{2}\right]\right]
\end{equation}

\noindent where b is a constant. Remarkably properties 1, 2, 3,and
5 are still true for this nonlinear form, but property 4 (separability)
fails. The contribution to the Lamb shift was also estimated for this
nonlinear form {[}11{]}. A difficulty for this type of model was pointed
out in {[}12{]} where the spectral absorption of a Hydrogen atom was
found not to have sharp line structure. 

\section{Conclusion}

The possibility of a logarithmic term in Schrödinger's equation with
a positive sign, opposite to that chosen by Bialynicki-Birula and
Micielski should not be ruled out as unphysical when considering further
experiments to look for evidence of a nonlinear term.

\section*{References}

\noindent {[}1{]} I. Bialynicki-Birula and J. Mycielski, Annals of
Physics 100, (1976) 62. 

\noindent {[}2{]} M. Davidson, Physica, 96A , (1979) 465. 

\noindent {[}3{]} M. Davidson, J. Math. Phys., 20 , (1979) 1865. 

\noindent {[}4{]} S. Weinberg, Phys. Rev. Lett. 62, (1989) 485. 

\noindent {[}5{]} A. Shimony, Phys. Rev. A 20, (1979) 394. 

\noindent {[}6{]} C. G. Shull, D. K. Atwood, J. Arthur, and M. A.
Horne, Phys. Rev. Lett. 44, (1980) 765. 

\noindent {[}7{]} C. G. Shull, A. Zeilinger, G. L. Squires, M. A.
Horne, D. K. Atwood, and J. Arthur, Phys. Rev. Lett.. 44, (1980) 1715. 

\noindent {[}8{]} R. Gahler, A. G. Klein, and A. Zeilinger, Phys.
Rev. A 23, (1981) 1611. 

\noindent {[}9{]} N. Gisin, Phys. Lett. A 143, (1990) 1. 

\noindent {[}10{]} N. Gisin, Helv. Phys. Acta 62, (1989) 363. 

\noindent {[}11{]} G. Cavalleri, and G. Mauri, Phys. Rev. B41, (1990)
6751; G. Cavalleri, and A. Zecca, Phys. Rev. B43, (1991) 3223. 

\noindent {[}12{]} A. Denis, in New Frontiers in Quantum Electrodynamics
and Quantum Optics, edited by A. O. Barut, (Plenum Press, New York)
1990, pp. 421-426.
\end{document}